\documentclass[prl,twocolumn]{revtex4}
\usepackage{graphicx}
\usepackage{color}
\usepackage{amsmath}
\usepackage{amssymb}
\usepackage{latexsym}
\usepackage{psfrag}

\begin{document}

\title{Spin polarization in open quantum dots}
\author{M. Evaldsson and I. V. Zozoulenko}
\affiliation{Department of Science and Technology (ITN), Link\"{o}ping University,
601 74 Norrk\"{o}ping, Sweden}
\date{\today}

\begin{abstract}
We investigate coherent transport through open lateral quantum dots using
recursive Green's function technique, incorporating exchange-correlation
effects within the Density Functional Theory (DFT) in the local spin-density approximation (LSDA). At low electron
densities the current is spin-polarized and electron density in the dot
shows a strong spin polarization. As the electron density increases the spin
polarization in the dot gradually diminishes. These findings are consistent
with available experimental observations. Results of our DFT-based modeling
indicate that utilization of the simplified approaches that use
phenomenological parameters and/or model Hamiltonians might not be always
reliable for theoretical predictions as well as interpretations of the
experiments.
\end{abstract}

\maketitle



\textit{Introduction}. A detailed understanding of spin-related phenomena in
quantum systems is necessary for future spintronics applications such as
spin filtering devices\cite{Slobodskyy}, spin-FETs \cite{DattaBiswajit},
nonvolatile computer memories \cite{Daughton}, etc. Semiconductor quantum
wires, dots and antidots defined in a two-dimensional electron gas (2DEG)
represent promising systems for the implementation of quantum spintronic
devices \cite{SpintronicDevices}. In this context, a topical question is
whether the spin degeneracy in these structures can be lifted.

Experimental studies indicate the existence of a spontaneous spin
polarization at low densities in the 2DEG\cite{Ghosh}. The spontaneous spin
polarization was suggested as the origin of \textquotedblleft
0.7-structure\textquotedblright\ in the conductance of a quantum points
contact (QPC) \cite{Thomas}. Concerning spin polarization in open quantum
dots, i.e., dots with strong coupling to leads as opposed to the Coulomb
blockade regime, the existing experiments show conflicting findings. A
statistical analysis of conductance fluctuations \cite{Marcus} indicated a
spin degeneracy at low magnetic fields. The similar conclusion follows from
the results of Folk \textit{et al}.\cite{Folk} who experimentally
demonstrated the operation of an open dot as a spin filter. In contrast,
low-field magnetoconductance of small dots \cite{Evaldsson} showed a
pronounced peak splitting that was taken as a signature of the spin
polarization. In contrast to QPCs, where theoretical investigations of the
spin polarization have been a subject of lively discussions during recent
years\cite{QPC1, QPC3, Hirose, Havu2004, Hirose2003, Matveev}, theoretical description of
spin-polarization effects in open quantum dots has received far less
attention. The main purpose of this paper is to provide such the description.

Modelling transport through quantum dots can be done from conceptually
different standpoints. For example, charging and coupling constants might be
taken as phenomenological parameters of the theory within a model
Hamiltonian \cite{Bulka, Evaldsson}. It is however not always evident
whether such a simplified description is sufficient to capture the essential
physics, and it is not always straightforward to relate quantitatively the
above parameters to the physical processes they represent in the real
system. Another common approach is to approximate the open system at hand
(which is described by the extended (scattering) states), by a corresponding
closed system, and then to calculate a self-consistent potential of this
system. The calculated potential is then used to obtain the transmission
coefficient on the basis of e.g.\ transfer matrix technique or related
methods\cite{Wang, QPC2}. This approach avoids ambiguity related to the
choice of a model Hamiltonian and phenomenological constants but depends on
the crucial (and not obvious) assumption that there is no qualitative
difference between many-electron effective potentials for open and closed
systems.

In the present paper we present an approach based on full quantum mechanical
many-electron magnetotransport calculations for \textit{open} system. Using
the recursive Green's function technique \cite{Sols} we compute the
scattering solutions of the two-dimensional Schr\"{o}dinger equation in a
magnetic field. Following the parametrization for the exchange and
correlation energy functionals of Tanatar and Cerperly \cite{TC}, the
electron-electron interaction is incorporated within the Density Functional
Theory (DFT) in the Local Spin Density Approximation (LSDA). The choice of
DFT+LSDA for the description of many-electron effects is motivated, on one
hand, by its efficiency in practical implementation within a standard
Kohn-Sham formalism\cite{ParrYang}, and, on the other hand, by the exellent
agreement between the DFT and exact diagonalization\cite{Ferconi} and
variational Monte-Carlo calculations\cite{Rasanen,QDOverview} performed for
few-electron quantum dots .

We stress that our approach involves no phenomenological parameters (like
coupling/charging constants), and that the self-consistent potential is
computed for the scattering states of an open system. As a note,
conceptually similar approaches have been used to model quantum transport in
molecular conductance and related molecular systems \cite{Taylor} as well as
to study quantum point contacts in zero magnetic field\cite{Havu2004, Hirose2003}. %
%

\begin{figure}[tbp]
\includegraphics{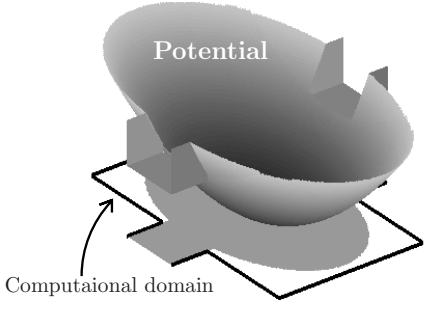}
\caption{External confinement potential $v_{ext}$ and the computational area
used for the dot. The strength of the confinement is $\hbar \protect\omega %
_{x}=0.2meV,\hbar \protect\omega _{y}=0.15meV$, the area size is 360$\times $%
210nm and the lead width is 80nm. Shaded area in the computational domain
indicates schematically the region where the confining potential lies below
the Fermi level.}\label{fig:external_potential}
\end{figure}

\textit{Model. }We consider a two dimensional (2D) lateral quantum dot
defined by a top gate in a rectangular geometry. The gates induce an
external electrostatic confinement described by a harmonic potential \cite%
{Kumar,QDOverview} %
\begin{equation}
v_{ext}(\mathbf{r})=\frac{1}{2}m^{\ast }\left( \omega _{x}^{2}x^{2}+\omega
_{y}^{2}y^{2}\right).   \label{eq:externalPotential}
\end{equation}%
The dot is connected to electron reservoirs by two semi-infinite leads, Fig.~\ref{fig:external_potential}.
The total confining potential includes also a contribution from a positive
background due to a uniform layer of donors \cite{Havu2004}. By changing the
donor concentration $\rho _{+}$, which contributes to the classical Hartree
potential (Eq.\ \ref{eq:hartreePotential}), we can effectively alter the
electron density in the dot. In this manner, the transport characteristics
of dots containing different electron densities can be investigated. In
the present paper we present a case study of the magnetotransport in a dot with
\textquotedblleft low\textquotedblright , \textquotedblleft
intermediate\textquotedblright\ and \textquotedblleft
high\textquotedblright\ electron densities. This corresponds to a donor concentration $\rho
_{+}/10^{14}\text{m}^{-2}=2,8,20$ (resulting in similar average electron densities, $\bar{n}_s$). The interaction parameter $r_s=1/a^*_B\sqrt{\pi\bar{n}_s}$, where $a^*_B$ is the effective Bohr radius, is 3.75, 2.00 and 1.28 respectively. 


The dot is described by the effective Hamiltonian in a perpendicular
magnetic field, $\mathbf{B_{\bot }}=B\mathbf{\hat{z}}$,
\begin{equation}
\hat{H}_{QD}^{\sigma }=H_{0}+v_{eff}^{\sigma }(\mathbf{r})+g\mu _{B}B\sigma ,
\label{eq:dotHamiltonian}
\end{equation}%
where $H_{0}$ is
\begin{equation}
H_{0}=-\frac{\hbar ^{2}}{2m^{\ast }}\Bigg\{\left( \frac{\partial }{\partial x%
}-\frac{eiBy}{\hbar }\right) ^{2}+\frac{\partial ^{2}}{\partial ^{2}y}\Bigg\}%
.  \label{eq:H0}
\end{equation}%
In Eq. (\ref{eq:H0}) the Landau gauge is chosen, $\mathbf{A}=(-By,0,0)$; $%
m^{\ast }$ is the effective mass in GaAs ($=0.067m_{e}$). The last term in
Eq. (\ref{eq:dotHamiltonian}) accounts for Zeeman energy where $\mu
_{B}=e\hbar /2m_{e}$ is the Bohr magneton, $\sigma =\pm
{\frac12}%
$ describes spin-up and spin-down states, $\uparrow $, $\downarrow $, and $%
g $ factor of GaAs is $g=-0.44.$

The effective potential, $v_{eff}^{\sigma }($\textbf{r}$)$, follows from the
Kohn-Sham theory \cite{QDOverview, ParrYang}, and consists of three parts,
\begin{equation}
v_{eff}^{\sigma }(\mathbf{r})=v_{H}(\mathbf{r})+v_{xc}^{\sigma }(\mathbf{r}%
)+v_{ext}(\mathbf{r}).  \label{eq:effective_potential}
\end{equation}%
$v_{H}($\textbf{r}$)$ is the classical Hartree potential due to the electron
density $n($\textbf{r}$)=\sum_{\sigma }n^{\sigma }($\textbf{r}$)$, and the
layer of positive donors, $\rho _{+}$
\begin{equation}
v_{H}(\mathbf{r})=\frac{e^{2}}{4\pi \varepsilon_0\varepsilon_r }\int d%
\mathbf{r}^{\prime }\frac{n(\mathbf{r}^{\prime })-\rho _{+}}{|\mathbf{r}-%
\mathbf{r}^{\prime }|}.  \label{eq:hartreePotential}
\end{equation}%
The exchange and correlation potential $v_{xc}($\textbf{r}$)=v_{x}($\textbf{r%
}$)+v_{c}($\textbf{r}$)$ in the local spin density approximation reads \cite%
{ParrYang}, 
\begin{equation}
v_{xc}^{\sigma }=\frac{d}{dn^{\sigma }}\Big(n^{\sigma }\epsilon
_{xc}(n,\zeta (\mathbf{r}))\Big)  \label{eq:LDA}
\end{equation}%
where $\zeta =\frac{n^{\uparrow }-n^{\downarrow }}{n^{\uparrow
}+n^{\downarrow }}$ is the spin-polarization. For $\epsilon _{xc}$ we have
used the parametrization of Tanatar and Cerperly \cite{TC}; the explicit
expressions for $v_{x}($\textbf{r}$)$ and $v_{c}($\textbf{r}$)$ can be found
in [\onlinecite{Macucci1993}].

%

The leads are modeled as hard wall potentials where the magnetic field is
restricted to zero. This is fully justified in the field interval under
consideration ($B\lesssim 0.5$ T), because the distortion of the wave
function in the leads due to the effect of magnetic field is negligible. All
the results presented here correspond to one propagating mode in the leads.

%
%

\textit{Calculations.} In order to compute the electron density and the
conductance of the dot we use the recursive Green's function technique\cite%
{Sols}. We discretize Eq.\ (\ref{eq:dotHamiltonian}) and introduce the
tight-binding Hamiltonian (with the lattice constant of $a=$ 10 nm), where
the perpendicular magnetic field is included by Peierl's substitution \cite%
{Datta}. Introducing the retarded Green's function,
\begin{equation}
\mathcal{G}^{\sigma }=(E-H^{\sigma }+i\epsilon )^{-1}
\label{eq:greensfunction}
\end{equation}%
we compute the density at site \textbf{r} through \cite{Datta}
\begin{equation}
n^{\sigma }(\mathbf{r})=-\frac{1}{\pi }\int_{-\infty }^{\infty }\mathrm{Im}[%
\mathcal{G}^{\sigma }(\mathbf{r},\mathbf{r},E)]f(E-E_{F})dE,
\label{eq:findDensity}
\end{equation}%
where $f$ is the Fermi-Dirac distribution and $E_{F}$ is the Fermi energy.
Equations (\ref{eq:dotHamiltonian})--(\ref%
{eq:findDensity}) are solved self-consistently for the spin up/down
densities. Note that a direct integration along the real axis in Eq.\ (\ref%
{eq:findDensity}) is rather ineffective as its numerical accuracy is not
sufficient to achieve convergence of the self-consistent electron density.
Because of this, we transform integration contour into the complex plane $%
\Im (E)>0,$ where the Green's function is much more smoother. This also
allows us to include bound states in the dot, with energies below any
propagating mode in the leads. We would like to stress that the Hamiltonian $%
\hat{H}^{\sigma }$ in Eq. (\ref{eq:greensfunction}) (and thus the
corresponding Green's function $\mathcal{G}^{\sigma }$) correspond to
\textit{an open }system representing the quantum dots and semi-infinite
leads. Using the converged, self-consistent potential, the transmission
coefficient $T_{\sigma }$ is computed from the Green's function between the
two leads, $T_{\sigma }(E)\propto \sum_{m,n}\langle \psi _{n}|\mathcal{G}%
^{\sigma }($\textbf{r}$_{L_{R}},$\textbf{r}$_{L_{L}},E)|\psi _{m}\rangle $,
where \textbf{r}$_{L_{R,L}}$ denotes the coordinate at the right
respectively left lead, and $\psi _{n,m}$ the wave functions in the leads%
\cite{Datta}. Finally, the conductance $G=\sum_{\sigma }G_{\sigma }$ is
found from the Landauer formula, $G_{\sigma }=-\frac{e^{2}}{h}\int_{-\infty
}^{\infty }T_{\sigma }(E)\frac{\partial f(E-E_{F})}{\partial E}dE.$

\textit{Results. }To outline the role of exchange and correlation
interactions we first study the magnetotransport in a quantum dot within the
Hartree approximation (i.e., when $v_{xc}(\mathbf{r})$ is not included in
the effective potential (\ref{eq:effective_potential})). Figure~\ref%
{fig:hartreePotentials}(a),(b) show the electron densities $n^{\uparrow }(%
\mathbf{r})$, $n^{\downarrow }(\mathbf{r})$ and the average spin-up/down density, $\bar{n}_s^{\uparrow }$/ $\bar{n}_s^{\downarrow }$, in the low density quantum
dot. As expected, the electron densities and the effective potential are
practically the same for spin-up/down electrons. Although the Zeeman
term lifts the degeneracy between the electrons of different spins, the difference between spin-up/down electrons is very small ($\vert\bar{n}_s^{\downarrow}/\bar{n}_s^{\uparrow}\vert\sim1.01$).
\begin{figure}[tbp]
{\includegraphics[width=0.45\textwidth]{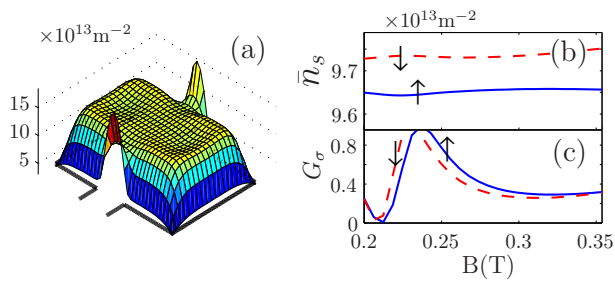}}
\caption{(color online) (a) The density for the spin-down electrons $%
n^{\downarrow }(\mathbf{r})$ in the low density dot calculated within the Hartree
approximation (note that $n^{\downarrow }(\mathbf{r})$ and $n^{\uparrow }(%
\mathbf{r})$ are indistinguishable on the scale of figure). (b) Average electron density for spin-down and spin-up electrons in the dot, $\bar{n}_s^{\uparrow },\bar{n}_s^{\downarrow }$
vs. magnetic field. (c) Conductance (in units of $e^2/h$) through the dot vs. magnetic field. Temperature T=2K.}
\label{fig:hartreePotentials}
\end{figure}
%
\begin{figure}[tbp]
\resizebox{\linewidth}{!}{\includegraphics{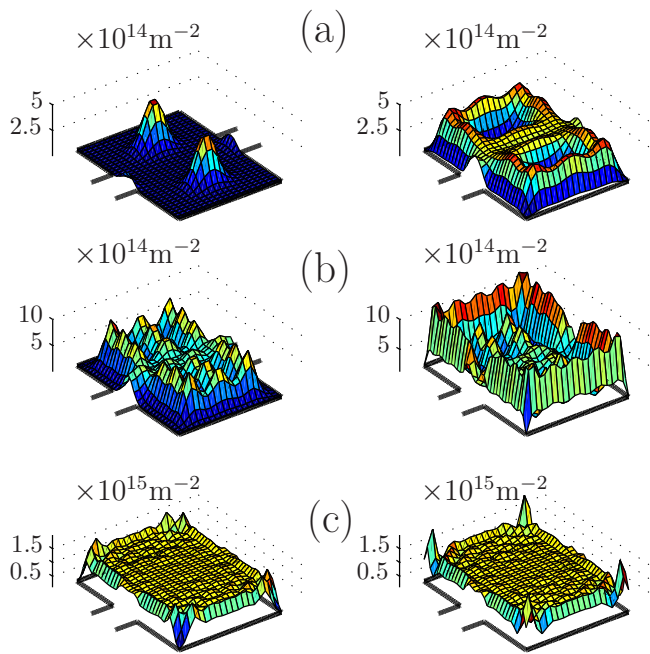}}
\caption{(color online) The density for the spin-up and spin-down electrons $%
n^{\uparrow }(\mathbf{r}),n^{\downarrow }(\mathbf{r})$ (left and right
panels respectively) in a (a) low, (b) intermediate and (c) high density dot
calculated within LSDA. T=2K.}
\label{fig:dftPotentials}
\end{figure}

\begin{figure}[tbp]
\includegraphics{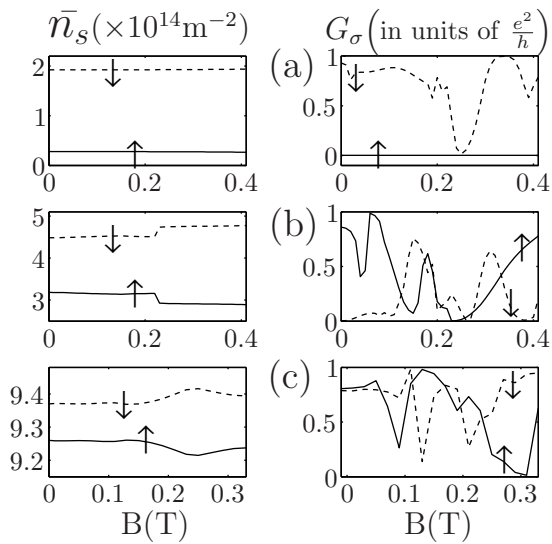}
\caption{Average electron density, $\bar{n}_s^{\uparrow }$, $\bar{n}_s^{\downarrow }$, in the
dot (left panel) and conductance (right panel) vs.\ magnetic
field for spin-up and spin-down electrons calculated within LSDA for (a)
low, (b) intermediate, (c) high density dots. $B=0$T in (a), (b), and $B=0.01$T
in (c). T=2K.}
\label{fig:trans_elan}
\end{figure}

Figures~\ref{fig:dftPotentials}(a) and \ref{fig:trans_elan}(a) show the
electron densities $n^{\uparrow }(\mathbf{r})$, $n^{\downarrow }(\mathbf{r}),$
the average electron densities, $\bar{n}_s^{\uparrow }$, $\bar{n}_s^{\downarrow }$, and the conductance
for the small quantum dot calculated in the framework of LSDA, where the
effective potential is given by Eq.\ (\ref{eq:effective_potential}). The
difference with the Hartree approximation (Fig.~\ref{fig:hartreePotentials})
is striking. The exchange interaction causes a significant spin polarization
with $\left\vert \bar{n}_s^{\downarrow }/\bar{n}_s^{\uparrow }\right\vert \thicksim 3$. Note
that spin-up and spin-down electrons are spatially separated and localized
in different parts of the dot. It is interesting to note that one of the
spin channels is completely suppressed such the conductance is dominated by
the spin-down electrons. This findings indicate that small quantum dots can
be used for injection of the spin-polarized current. It is also important to
stress that the spin polarization persists for $B=0,$ which clearly points
out to the exchange (not Zeeman) interaction as the origin of the effect.

\bigskip With the increase of the electron density the spin polarization in
the dots decreases. This trend is illustrated in Fig.~\ref{fig:trans_elan}~(b)
where the spin polarization in the intermediate density dot is decreased to $%
\left\vert \bar{n}_s^{\downarrow }/\bar{n}_s^{\uparrow }\right\vert \thicksim 1.2,$ and for
the high density dot is practically disappeared $\left\vert \bar{n}_s^{\downarrow
}/\bar{n}_s^{\uparrow }\right\vert \thicksim 1.01$. This behavior is also clearly
manifested in the distribution of the electron densities, Fig.~\ref%
{fig:dftPotentials}. In the intermediate density dot the spin-up/down densities $%
n^{\uparrow }(\mathbf{r})$, $n^{\downarrow }(\mathbf{r})$ are no longer as
distinctly separated as in the small one, and in the large dot $n^{\uparrow
}(\mathbf{r})$, $n^{\downarrow }(\mathbf{r})$ are, except at the edges,
practically the same throughout the dot. The peculiarities at the edges in %
\ref{fig:dftPotentials}(c) is due to the finite size of our computational
domain; as the density and Coulomb interaction increase the electrons are
squeezed towards the walls of our confinement. Different electron densities $%
n^{\uparrow }(\mathbf{r})$, $n^{\downarrow }(\mathbf{r})$ lead to different
potentials, which, in turn, result in the difference in the conductance for
two spin channels, $G_{\downarrow }$, $G_{\uparrow }$ (Fig.~\ref{fig:trans_elan}%
, left panel).

\bigskip The degree of spin polarization in open dots was probed by Folk
\textit{et al.} [\onlinecite{Marcus}] who concluded that the dots were spin
degenerate at low field. At the same time, the results of Ref.\ [%
\onlinecite{Evaldsson}] suggested that the spin degeneracy in open dots is
lifted. Our findings reported above seem to reconcile these apparently
different experimental observations. Indeed, the dots studied in Ref.\ [%
\onlinecite{Marcus}] had relatively large electron density ($\bar{n}_s$ $\sim 2\times10^{15}\mathrm{m}^{-2}$),
whereas the dots in Ref.\ [\onlinecite{Evaldsson}] had low electron density, ($\bar{n}_s<7\times10^{14}\mathrm{m}^{-2}$), such that the observed features \cite{Marcus,Evaldsson} are
consistent with our findings that spin polarization diminishes as the
electron density in the dot increases.

During recent years a significant attention has been devoted to the study of
the so-called \textquotedblleft 0.7-anomality\textquotedblright\ observed in
quantum point contacts (QPCs) \cite{Thomas}. Though still a highly debated
issue, this feature has been attributed by many researches to the effect of
spin polarization that occurs at low densities when the exchange energy
becomes more prominent compared to the kinetic energy \cite{Havu2004, QPC1,
QPC2}. Our present calculations show similar trends in the density dependent
spin-polarization in open quantum dots, indicating the similar origin of the
effect for these systems. Note that some recent studies question the
existence of the spin-polarization in QPCs and attribute the
\textquotedblleft 0.7-anomality\textquotedblright\ to different mechanisms,
in particular, to the formation of one-dimension (1D) Wigner crystal\cite%
{Matveev}. Such the interpretation relies on the notion of 1D interacting
system. In contrast, our findings strongly indicate that the
spin-polarization at low electron densities is a generic feature of both
quasi-1D and quasi-0D systems (i.e.\ QPC and open dots). Because of this, we
hope that our results will stimulate further experimental studies of the
open quantum dots which might help to settle still unresolved issue
concerning the spin polarization in the above systems.

In our previous work \cite{Evaldsson}, we investigated transport in similar
dots using the mean-field Hubbard hamiltonian, $H^{\sigma
}=H_{0}+U\,n_{\uparrow }(\mathbf{r})n_{\downarrow }(\mathbf{r}),$ where
on-site Hubbard constant $U$ describes the Coulomb interaction between
electrons of opposite spins. Within this model we found that spin
polarization in the dot takes place only when the Fermi energy hits the
resonant states weakly coupled to the leads. In contrast, the present
calculations based on the LSDA shows qualitatively different behavior, when
the spin polarization is nearly constant in the broad interval of magnetic
field. This difference strongly indicates that utilization of the simplified
approaches that use phenomenological parameters and/or model Hamiltonians
might not be always reliable for theoretical predictions as well as
interpretations of the experiments.

%

\textit{Conclusion.} We have performed full quantum mechanical calculations
of the magnetotransport in open quantum dots where the electron- and spin
interactions have been incorporating within the local spin density
approximation. At low electron densities the current is spin-polarized and
electron density in the dot shows a strong spin polarization. As the
electron density increases the spin polarization in the dot gradually
diminishes. These findings are consistent with the existing experimental
observations.

%
%
\textit{Acknowledgements. }Financial support from the National Graduate
School in Scientific Computing (M.E.) and the Swedish Research Council (M.E.
and I.V.Z) is acknowledged. We are thankful to Andrew Sachrajda for
discussions and critical reading of the manuscript.


\vspace{0.25cm}

\begin{thebibliography}{99}
\bibitem{Slobodskyy} A. Slobodskyy, C. Gould, T. Slobodskyy, C. R. Becker,
G. Schmidt, and L. W. Molenkamp, Phys. Rev. Lett. \textbf{90}, 246601, (2003)

\bibitem{DattaBiswajit} S. Datta and B. Das, Appl. Phys. Lett. \textbf{56},
665 (1990)

\bibitem{Daughton} J. M. Daughton, Thin Solid Films \textbf{216}, 162 (1992)

\bibitem{SpintronicDevices} G. Burkard, D. Loss, and D. P. DiVincenzo, Phys.
Rev. B \textbf{59}, 2070 (1999); P. Recher, E. V. Sukhorukov, and D. Loss,
Phys. Rev. Lett. \textbf{85}, 1962 (2000); I. V. Zozoulenko and M.
Evaldsson, Appl. Phys. Lett. \textbf{85}, 3136 (2004).


\bibitem{Ghosh} A. Ghosh, C. J. B. Ford, M. Pepper, H. E. Beere, and D. A.
Ritchie, Phys. Rev. Lett. \textbf{92}, 116601 (2004)

\bibitem{Thomas} K. J. Thomas, J. T. Nicholls, M. Y. Simmons, M. Pepper, D.
R. Mace, and D. A. Ritchie , Phys. Rev. Lett. \textbf{77}, 135 (1996)

\bibitem{Marcus} J. A. Folk, S. R. Patel, K. M. Birnbaum, C. M. Marcus, C.
I. Duru\"{o}z and J. S. Harris, Jr., Phys. Rev. Lett. \textbf{86}, 2102
(2001).

\bibitem{Folk} J. A. Folk, R. M. Potok, C. M. Marcus, and V. Umansky,
Science \textbf{299}, 679 (2003).

\bibitem{Evaldsson} M. Evaldsson, I. V. Zozoulenko, M. Ciorga, P. Zawadzki
and A. S. Sachrajda, Europhys. Lett. \textbf{68}, 261 (2004)

\bibitem{QPC3} V. V. Flambaum and M. Yu. Kuchiev, Phys. Rev. B \textbf{61},
R7869 (2000)


\bibitem{QPC1} C.-K. Wang and K.-F. Berggren, Phys. Rev. B \textbf{54}, R14
257 (1996)

\bibitem{Hirose} K. Hirose and N. S. Wingreen, Phys. Rev. B \textbf{64},
073305 (2001)

\bibitem{Havu2004}P. Havu, V. Havu, M. J. Puska, and R. M.
Nieminen, Phys. Rev. B \textbf{69}, 115325 (2004)

\bibitem{Hirose2003} K. Hirose, Y. Meir and N. S. Wingreen, Phys. Rev. Lett.
\textbf{90}, 026804 (2003)

\bibitem{Matveev} K. A. Matveev, Phys. Rev. B \textbf{70}, 245319 (2004).

\bibitem{Bulka} A. Groshev, T. Ivanov, and V. Valtchinov, Phys. Rev. Lett.
\textbf{66}, 1082 (1991); C. Niu, L.-J. Liu, and T.H. Lin, Phys. Rev. B
\textbf{51}, 5130 (1995); B. R. Bu\l ka and P. Stefa\'nski, Phys. Rev. Lett.
\textbf{86}, 5128 (2001)

\bibitem{Wang} D. Jovanovic and J.-P. Leburton, Phys. Rev. B \textbf{49}, 7474
(1994); Y. Wang, J. Wang, H. Guo and E. Zaremba , Phys. Rev. B \textbf{52},
2738 (1995)


\bibitem{QPC2} K.-F. Berggren and I. I. Yakimenko, Phys. Rev. B \textbf{66},
085323 (2002);

\bibitem{Sols} F. Sols, M. Macucci, U. Ravaioli, and K. Hess, J. Appl. Phys.
\textbf{66}, 3892 (1989)

\bibitem{TC} B. Tanatar and D. M. Ceperley, Phys. Rev. B \textbf{39}, 5005,
(1989)

\bibitem{ParrYang} R. G. Parr and W. Yang, \textit{Density-Functional Theory
of Atoms and Molecules}, (Oxford Science Publications, Oxford, 1989).

\bibitem{Ferconi} M. Ferconi and G. Vignale, Phys. Rev. B \textbf{50},
R14722 (1994).

\bibitem{Rasanen} E. R\"{a}s\"{a}nen, H. Saarikoski, V. N. Stavrou, A.
Harju, M. J. Puska, and R. M. Nieminen, Phys. Rev B \textbf{67}, 235307
(2003).

\bibitem{QDOverview} S. M. Reimann and M. Manninen, Rev. Mod. Phys. \textbf{%
74}, 1283 (2002)

\bibitem{Taylor} P. S. Damle, A. W. Ghosh, and S. Datta, Phys. Rev. B
\textbf{64}, 201403(R) (2001); J. Taylor, H. Guo and J. Wang, Phys. Rev. B
\textbf{63}, 245407 (2001)


\bibitem{Kumar} A. Kumar, S. E. Laux and F. Stern, Phys. Rev. B, \textbf{42}%
, 5166 (1990)

\bibitem{Macucci1993} M. Macucci, Karl Hess and G.J. Iafrate, Phys. Rev. B
\textbf{48}, 17354 (1993)

\bibitem{Datta} S. Datta, \textit{Electronic Transport in Mesoscopic Systems}%
, (Cambridge University Press, Cambridge, 1997)
\end{thebibliography}
\end{document}